\documentclass[12pt,journal,draftclsnofoot,oneside,onecolumn]{IEEEtran}

\usepackage{color,soul}
\usepackage[cmex10]{amsmath}
\usepackage{amssymb}
\usepackage{epsfig}
\usepackage{stfloats}
\usepackage{cite}
\usepackage{multirow}

\begin{document}

\title{Secure Non-Orthogonal Multiple Access: An Interference Engineering Perspective}
\author{Lu Lv, Hai Jiang, Zhiguo Ding, Qiang Ye, Naofal Al-Dhahir, and Jian Chen
\thanks{\it Lu Lv is with Xidian University, and also with Southeast University; Hai Jiang is with University of Alberta; Zhiguo Ding is with The University of Manchester; Qiang Ye is with Dalhousie University; Naofal Al-Dhahir is with The University of Texas at Dallas; Jian Chen (corresponding author) is with Xidian University.}
}
\markboth{IEEE Network (Accepted from Open Call)}{}
\maketitle

\begin{abstract}
  Non-orthogonal multiple access (NOMA) is an efficient approach that can improve spectrum utilization and support massive connectivity for next-generation wireless networks. However, over a wireless channel, the superimposed NOMA signals are highly susceptible to eavesdropping, potentially leading to severe leakage of confidential information. In this article, we unleash the potential of network interference and exploit it constructively to enhance physical-layer security in NOMA networks. Particularly, three different types of network interference, including artificial noise, specifically-designed jamming signals, and inter-user interference, are well engineered to intentionally reduce information leakage while mitigating the effect on signal reception quality of legitimate users, thereby significantly enhancing the transmission security of NOMA. Furthermore, we propose interference engineering strategies for more advanced full-duplex NOMA, intelligent reflecting surface NOMA, cognitive radio NOMA, and multi-cell NOMA networks, and discuss several open research problems and challenges, which could inspire innovative interference engineering designs for secure NOMA communications.
\end{abstract}

\IEEEpeerreviewmaketitle

\section{Introduction}

It is very challenging for current wireless networks to serve the ever-increasing Internet-of-Things (IoT) applications over the scarce spectrum using orthogonal resource allocation. Non-orthogonal multiple access (NOMA) is a promising technique to break the orthogonality constraint for resource allocation \cite{DaiLinglong_CM2015}. Using joint superposition coding (SC) and successive interference cancellation (SIC), NOMA is able to serve multiple users in the same time-frequency resource, thus improving spectral efficiency and achieving massive connectivity. In addition, NOMA has the advantages of better fairness and lower latency compared to the conventional orthogonal multiple access (OMA), and is considered as a competitive candidate multiple access technique for next-generation wireless networks.

\begin{figure}[t]
  \centering
  \includegraphics[width=4.5in]{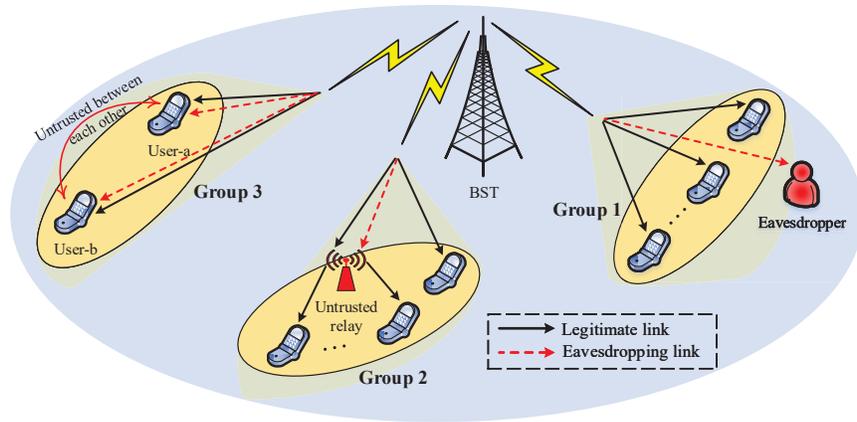}
  \caption{Typical passive eavesdropping attacks in NOMA networks.} \label{fig:eavesdropping}\vspace{-3mm}
\end{figure}

Despite the aforementioned advantages, the broadcast nature of wireless channels poses new security challenges to NOMA networks, namely, the transmitted NOMA signals are vulnerable to passive eavesdropping attacks (where an eavesdropper remains silent while trying to decipher the confidential information based on its received NOMA signals), as illustrated in Fig.~\ref{fig:eavesdropping}. Here, a base-station (BST) uses three orthogonal beams to connect three groups of users, and within each beam NOMA is applied to serve multiple users. In group~1, there is an external eavesdropper, while in group 2, an untrusted relay is used (i.e., the relay employs the designated amplify-and-forward protocol but may also try to decode the confidential information of the users). After the eavesdropper or the untrusted relay receives the superimposed signal from the BST, they can adopt SIC to decode and intercept all users' information contained in the superimposed signal. In contrast, if OMA is used, by receiving a signal over a resource block, the eavesdropper or untrusted relay can only intercept one user's information. {\it Therefore, compared to OMA transmission, the use of NOMA triggers a more severe information leakage problem.} In group~3, there are two users served by NOMA, and each user may serve as an internal eavesdropper and take advantage of SIC to overhear the signals of the other user. For example, if user-a (the near user) is an eavesdropper, it can decode user-b (the far user)'s signal easily, as user-a is designed to decode user-b's signal first in SIC. If user-b is an eavesdropper, after retrieving its own signal, it tries to decode user-a's signal through SIC. {\it Thus, even without an external eavesdropper, the use of NOMA gives rise to internal privacy leakage.} As a result, how to secure wireless NOMA communications against external/internal eavesdropping attacks is a new and challenging issue to address.

In this context, physical-layer security has been recognized as an efficient means to enhance the transmission security of NOMA. Specifically, physical-layer security exploits the intrinsic properties of wireless media to improve the secrecy rate, i.e., the rate of the legitimate channel (from a legitimate transmitter to its legitimate receiver) minus that of the wiretap channel (from the legitimate transmitter to an eavesdropper) \cite{WuYongpeng_JSAC2018}. Among all physical-layer security solutions, the use of network interference is highly appealing to secure NOMA communications, due to the following motivations:

\begin{itemize}
  \item Intentionally generated interference, such as artificial noise (AN) and specifically-designed jamming signals (SJS) at the eavesdropper's side, can make the wiretap channel a degraded version of the legitimate channel, which helps achieve a positive secrecy rate, i.e., perfect security.

  \item Inter-user interference (IUI), which is inherent in NOMA due to the power-domain multiplexing, is traditionally treated as a deleterious factor to the legitimate users. However, the IUI also affects the eavesdropper and can be engineered towards a security advantage. By striking a balance between reducing the IUI at the legitimate users and amplifying the IUI at the eavesdropper, the secrecy rate can be increased significantly.
\end{itemize}

In this article, we unveil the beneficial role of network interference for physical-layer security and investigate interference engineering strategies for secure NOMA communications. We provide an overview of three existing solutions that intelligently engineer the AN, SJS, and IUI to improve the transmission security of NOMA against passive eavesdropping attacks. Subsequently, by integrating NOMA with advanced wireless communication concepts, we propose several interference engineering strategies to exploit more network interference resources, which are inherent in full-duplex NOMA, intelligent reflecting surface (IRS) NOMA, cognitive radio NOMA, and multi-cell NOMA networks. Some open research problems and challenges are also discussed.

\section{Secure NOMA Communications via Artificial Noise Engineering}
\label{sec:AN-engineering}

In some cases, the distance from an eavesdropper to a BST may be shorter than those from the NOMA users to the BST. Due to the strong channel gain of the eavesdropper, any signal decoded by the users can also be decoded by the eavesdropper. Hence, it is beneficial to generate random AN to impair the signal reception quality of the eavesdropper. Ideally, AN should be engineered in the null space of the legitimate channels while only the wiretap channel is affected, which can be achieved using secure beamforming and cooperative jamming shown as follows.

\begin{figure}[t]
  \centering
  \includegraphics[width=5.6in]{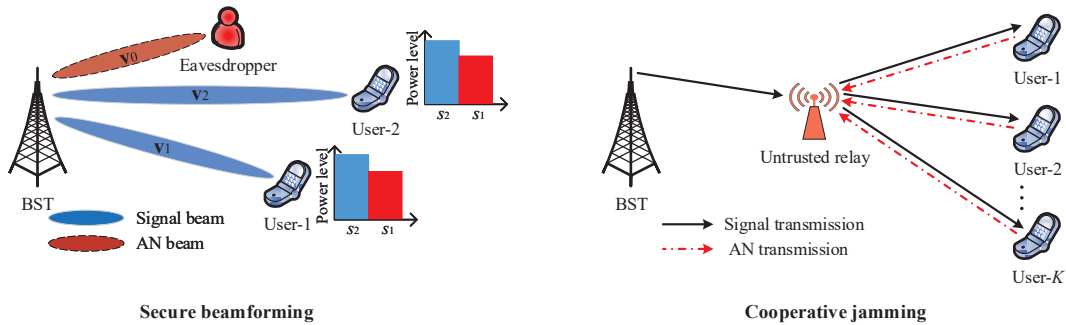}
  \caption{AN engineering for secure NOMA communications: secure beamforming and cooperative jamming.} \label{fig:an-exploitation}\vspace{-3mm}
\end{figure}

{\bf Secure Beamforming:} The basic idea of secure beamforming is to let a user simultaneously transmit its signal and AN by using multiple-antenna techniques. In the conventional OMA scenario with only a single user, the signal beam is designed based on maximum ratio transmission (MRT) to enhance the reception quality of the user, and the AN beam is designed based on the zero-forcing (ZF) metric to avert the negative effect of the AN on the user. However, the use of NOMA brings fundamental challenges to the design of secure beamforming \cite{Lu_TVT2018}.

One challenge is that employing MRT based signal beams for different users is no longer suitable for NOMA. This can be illustrated by using a typical two-user NOMA example shown on the left side of Fig.~\ref{fig:an-exploitation}, where the signals intended for user-1 (who has a strong channel gain) and user-2 (who has a weak channel gain) are denoted by $s_1$ and $s_2$, respectively. If MRT is applied, i.e., the BST independently precodes the signal beam $\mathbf{v}_1$ (for transmitting $s_1$) based on the channel gain of user-1 and the signal beam $\mathbf{v}_2$ (for transmitting $s_2$) based on the channel gain of user-2, the received signal strengths of $s_1$ and $s_2$ are maximized at user-1 and user-2, respectively. This may result in a SIC failure at user-1, due to the fact that the IUI caused by $s_1$ is also maximized when user-1 first decodes $s_2$ in SIC, and thus, the system performance is degraded significantly. To guarantee both reliability and secrecy in NOMA transmissions, the signal beams $\mathbf{v}_1$ and $\mathbf{v}_2$ should be jointly designed using a weighted combination of the channel gains of both users \cite{YanShihao_TWC2019}, where the weighting coefficients are optimized to satisfy that: 1) At user-1, the received signal powers of $s_1$ and $s_2$ are boosted simultaneously to guarantee user-1's perfect SIC and signal decoding, and 2) at user-2, the received signal power of $s_2$ is enhanced while the received interference power of $s_1$ is suppressed to benefit user-2's successful signal decoding. In addition, the AN beam $\mathbf{v}_0$ is engineered via ZF without affecting the legitimate channels of both users. As a result, by increasing the transmit power of the BST, the achievable rates of the users are improved while the achievable rate of the eavesdropper saturates, and perfect security for NOMA communications is achieved \cite{YanShihao_TWC2019}.

Another challenge is that transmitting the AN in the null space of all legitimate channels may be impossible for a large number of NOMA users, since in this case the BST does not have sufficient degrees of freedom (DoF) to perform ZF. To address this problem, one viable solution is to employ user grouping and scheduling, i.e., first dividing multiple users into several orthogonal groups with each group occupying a single resource block (i.e., a time slot), and then using NOMA to schedule the users in each group \cite{Yuanwei_TWC2017}. As such, the BST has enough DoF to perfectly null out the AN in the user group and make the AN solely affect the eavesdropper, thereby ensuring the security of NOMA.

Note that secure beamforming also performs well in the presence of multiple eavesdroppers. This is because AN is transmitted isotropically except towards the legitimate users in secure beamforming, and the performance of all the eavesdroppers will be degraded by the AN.

{\bf Cooperative Jamming:} In untrusted relay networks with no direct link between a BST and a user, cooperative jamming is an efficient means to prevent information from being leaked to the untrusted relay. In OMA, this strategy exploits the user as a helper in the first phase to send the AN to jam the relay when the BST is transmitting to the relay, and as a receiver in the second phase to retrieve the BST's signal after employing the self-AN cancellation. Therefore, only the reception quality of the relay is degraded by the AN, which is beneficial to improve the secrecy performance. As compared to the traditional OMA case, cooperative jamming in NOMA needs extra efforts in dealing with more sophisticated AN cancellation at the user side illustrated on the right side of Fig.~\ref{fig:an-exploitation}. Particularly, the BST applies NOMA to communicate with multiple users via an untrusted relay, where all the users assist the BST's secrecy transmission by sending AN in a collaborative manner to the relay \cite{Ahmed_TIFS2020}. To make the AN cancellation possible, each user has to know the AN not only from itself but also from the other users {\it a priori}. Therefore, the physical-layer key distribution mechanism can be performed, where all users share the seed information of the AN generation to facilitate the AN cancellation. Although using all the users for cooperative jamming can maximally degrade the reception quality of the relay, the involved physical-layer key distribution inevitably incurs a very high computational complexity and is not scalable as the number of NOMA users becomes large. To simplify the AN cancellation at the users, an alternative approach is to select one user to transmit the AN and let the other users receive and cache the AN in the first phase. Then, in the second phase, all the users can cancel the received AN completely since a copy of the AN is transmitted by the selected user and cached by the other users previously \cite{YangWeiwei_WCL2019}. In this way, secrecy against the untrusted relay is promised by cooperative jamming without coordination, thereby ensuring low complexity. On the other hand, a single user only has limited transmit power and DoF to jam the relay effectively, particularly when the relay has a powerful signal detection/cancellation ability (e.g., with multiple antennas). Multiuser cooperative jamming is a promising strategy to combat the powerful untrusted relay. Hence, a multiuser scheduling scheme has to be developed to optimize the trade-off between the secrecy performance and complexity.

Cooperative jamming also performs well when multiple untrusted relays are adopted to help forward information from the BST to the users. This is because the AN, which is transmitted by either all the users or one selected user, can degrade reception quality of all untrusted relays.

Several future research directions are discussed next. The aforementioned AN engineering strategies mainly focused on the communication security, but neglected the reliability performance. In fact, the security and reliability are coupled and affect each other. For instance, although the AN is helpful in confusing the eavesdropper, the AN generation consumes extra transmit power which leads to less transmit power available for signal transmission, and this degrades the reception quality of the users. Therefore, carefully engineering the AN to achieve a balanced security-reliability tradeoff is an open and interesting problem. Furthermore, in cooperative jamming with two-way untrusted relaying, the AN sent by the users cannot be canceled by the BST and affects the BST's signal reception. In this context, how to design secure and reliable cooperative jamming for two-way untrusted relaying is a challenging future research problem.

\section{Secure NOMA Communications via SJS Engineering}
\label{sec:SJS-engineering}

One unique feature of NOMA is that the receiver has the SIC capability, which provides an opportunity to use the SJS for intentionally degrading the eavesdropper's performance. Unlike the AN which has similar characteristics to the additive noise and cannot be decoded, the SJS is generated by using random codewords of a public codebook known by the users and the eavesdropper. The key issue is how to properly transmit the SJS, such that each user can decode and remove the SJS in the first step of SIC while the eavesdropper cannot do so, which is nontrivial. To resolve this issue, two promising strategies can be employed, namely, transmission rate control for SJS and decoding order design for SJS, as elaborated next.

\begin{figure}[t]
  \centering
  \includegraphics[width=5.6in]{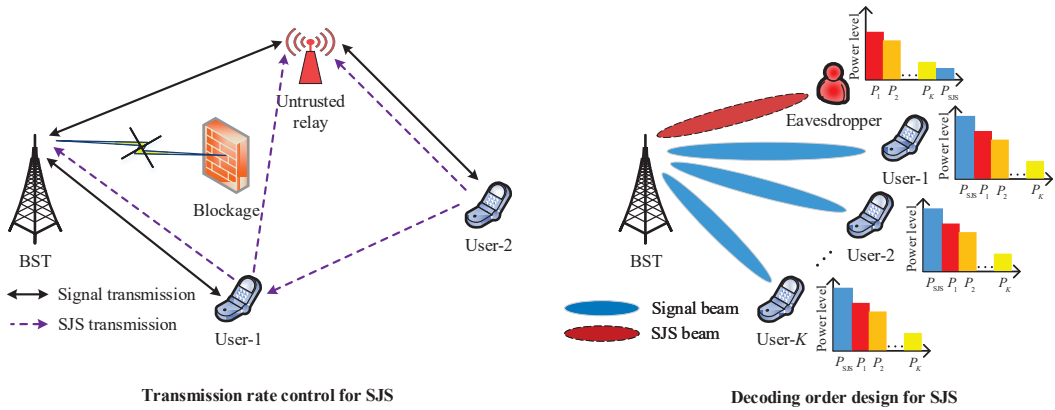}
  \caption{SJS engineering for secure NOMA communications: transmission rate control and decoding order design.} \label{fig:sjs-exploitation}\vspace{-3mm}
\end{figure}

\begin{figure}[t]
  \centering
  \includegraphics[width=3.6in]{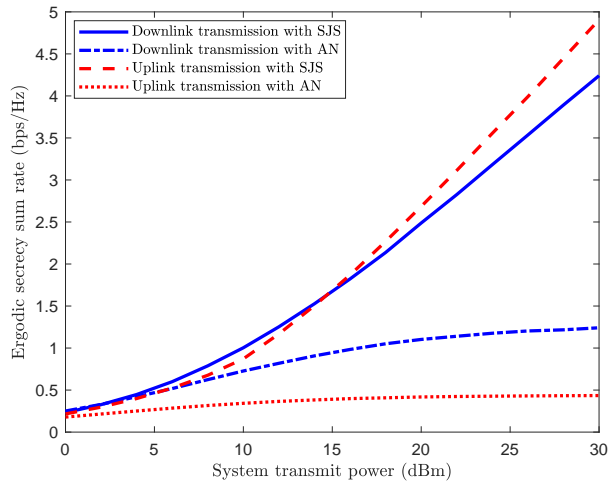}
  \caption{The ergodic secrecy sum rate performance. The average channel gains from the BST to relay, from the relay to user-2, from the BST to user-1, from user-1 to user-2, and from the relay to user-1 are set to 0.9, 0.9, 0.8, 1, and 0.5, respectively.} \label{fig:sjs-an}\vspace{-3mm}
\end{figure}

{\bf Transmission Rate Control for SJS:} We consider a NOMA-based coordinated direct and untrusted relay transmission (on the left side of Fig.~\ref{fig:sjs-exploitation}) to illustrate this strategy. The uplink transmission consists of two phases. In the first phase, when user-2 is transmitting, user-1 simultaneously transmits its signal and the SJS. Specifically, the transmission rate for the SJS is adapted according to the channel gain between user-1 and the BST, so that the BST can first decode and cancel the SJS by SIC without affecting its own signal decoding. The relay also tries to decode the SJS, however, the signals of both users will serve as interference to the SJS decoding, which results in a very low achievable rate smaller than the transmission rate of the SJS. Hence, the relay cannot decode the SJS. In the second phase, the relay forwards its received signals and user-1 transmits a new signal both to the BST. User-1's new signal is perfectly secured as the relay works in a half-duplex mode. The downlink transmission also comprises two phases. In the first phase, when the BST is transmitting, user-2 concurrently transmits the SJS. To ensure that user-1 can decode the SJS before decoding the signals of the BST, the transmission rate of the SJS is selected based on channel knowledge of the user-2$\leftrightarrow$user-1$\leftrightarrow$BST link, which is independent of that of the user-2$\leftrightarrow$relay$\leftrightarrow$BST link. Thus, it is very unlikely for the relay to decode and cancel the SJS. In the second phase, the relay forwards the received signal and the BST transmits a new signal to user-1. It is clear that in both the uplink and downlink, the SJS serves as useful jamming to degrade the wiretap channels of the relay, but does not affect the legitimate channels of the users/BST \cite{Lu_TCOM2020}, thereby improving the secrecy rate performance. In Fig.~\ref{fig:sjs-an}, the ergodic secrecy sum rate (ESSR) of this SJS engineering strategy is compared to that of the AN based strategy, i.e., user-1 and user-2 transmit AN in stead of the SJS in the uplink and downlink, respectively. It is observed that the SJS engineering strategy achieves a much higher ESSR than the AN based strategy in both the uplink and downlink, and the performance gap between the two strategies increases as the system transmit power increases. This reveals that the SJS engineering is helpful in safeguarding physical-layer security of NOMA, especially when the use of AN interferes with the signal reception quality of the legitimate users.

When multiple untrusted relays are used, this SJS engineering strategy can also be applied with minor modifications. In the first phase, the transmission rate of the SJS should be chosen such that the SJS can be decoded by user-1/BST but not by any of the untrusted relays, with more stringent transmission rate control. In the second phase, all the relays use distributed beamforming to forward signals.

{\bf Decoding Order Design for SJS:} As shown on the right side of Fig.~\ref{fig:sjs-exploitation}, the $M$-antenna BST serves $K$ NOMA users in the presence of an eavesdropper, where $M<K+1$. As discussed previously, even equipped with multiple antennas, secure beamforming via AN fails to achieve a favorable security performance due to the AN leakage issue. To overcome this difficulty, the BST can transmit the SJS and create distinguished decoding orders for the SJS at the users and the eavesdropper. Assume that the BST knows the channel state information (CSI) of the users and the eavesdropper. Then the BST can design its beamforming vectors to increase the received power of the SJS to the highest at the users while decreasing the received power of the SJS to the lowest at the eavesdropper \cite{NanZhao_TCOM2019}. As such, each user can first decode the SJS with SIC to mitigate its adverse effect on the users, but the eavesdropper fails to recover the SJS because its decoding order for the SJS is the last. Furthermore, by maximizing the transmit power of the SJS subject to the constraint on the achievable rates of the users, the eavesdropping capability is impaired as much as possible while the reliability performance of the users is guaranteed.

In the case of multiple eavesdroppers, the beamforming vectors of the BST should be designed to decrease the received power of the SJS to the lowest level at all the eavesdroppers, at the cost of more stringent power optimization.

Apart from the above progress, there are other interesting yet unsolved issues. For instance, instead of solely controlling the transmission rate or the decoding order of the SJS, a joint optimization of the transmission rate and the decoding order of the SJS and NOMA signals is desired to further enhance the communication security. In addition, secrecy energy efficiency, defined as the sum secrecy rate per Joule energy consumption, is an important measure to quantify the efficient use of the SJS for a tradeoff between energy savings and secrecy guarantees. Research on secrecy energy efficiency maximization for SJS engineering is still missing in the literature, which motivates future research works.

\section{Secure NOMA Communications via IUI Engineering}
\label{IUI-engineering}

Another unique feature of NOMA is the inherent IUI, which traditionally plays a harmful role in reducing the signal reception quality of the users. In fact, the IUI simultaneously affects the users and the eavesdropper. If appropriately engineered, the IUI can be beneficial to improve secrecy by serving as jamming signals to confuse the eavesdropper. This approach is cost-effective, in the sense that the secrecy is guaranteed not by constructing AN or SJS that consumes extra communication resources, but rather by reusing the IUI that already exists in NOMA networks. In general, the IUI can be engineered to achieve goals of hindering SIC at eavesdropper, impeding coherent detection at eavesdropper, and creating strong IUI at untrusted users, as discussed below.

{\bf Hindering SIC at Eavesdropper:} Indeed, secrecy can be enhanced by carefully harnessing the IUI to disable SIC at the eavesdropper. We explore this concept by investigating a NOMA network comprised of a $K$-antenna BST, $M$ users, and an eavesdropper. It is assumed that only $M_0$ ($<K$) users require secure transmission (in this case, the number of antennas at the BST is sufficient for ZF beamforming with respect to these users). The users that require secure transmission are called secure users (SUs) while the other users are called regular users (RUs), where the number of RUs is assumed, without loss of generality, to be more than that of SUs. A joint design of ZF beamforming, user clustering and scheduling can be applied to efficiently exploit the IUI for impeding the eavesdropper's SIC \cite{WangHuiming_JSTSP2019}, which is elaborated as follows. First, the BST groups all the users into $M_0$ clusters with one SU and one or more RUs in each cluster. After user clustering, the IUI exists in two forms: inter- and intra-cluster interference. Then, the BST designs its beams for each SU following the ZF principle, and schedules the RUs in each cluster to guarantee that the SU has the largest effective channel gain. In this way, each SU receives no inter-cluster interference and can subtract the intra-cluster interference by SIC. However, the eavesdropper does not know the ZF beams and the signal decoding order, and its SIC is prohibited. Hence, the eavesdropper is severely jammed by both inter- and intra-cluster interference. This strategy also performs well with multiple eavesdroppers, since the resultant IUI degrades the reception quality of all the eavesdroppers.

{\bf Impeding Coherent Detection at Eavesdropper:} When the eavesdropper has a large number of antennas, the AN approach may not work well since the eavesdropper can perform null-space receive beamforming to eliminate the AN. In this situation, the system can exploit the IUI to degrade the eavesdropper's detection performance \cite{FanYe_TVT2019}. Consider that a $K$-antenna BST communicates with two single-antenna users in the presence of an $N$-antenna eavesdropper with $K\ll N$. In each transmission, the BST splits each user's message into two parts. Through NOMA signaling, the first parts of the messages of the two users are superimposed into signal $s_1$, and the second parts of the messages of the two users are superimposed into signal $s_2$. After that, the BST linearly precodes $s_1$ and $s_2$ based on MRT and random beamforming (RB), respectively. Here, the RB vector is designed to guarantee that the users still experience a block fading channel while the eavesdropper undergoes an equivalent fast fading channel. As a result, each user can do coherent detection to reliably retrieve its desired signal by SIC, while the eavesdropper can only detect the signals non-coherently without SIC, which significantly degrades its bit-error-rate performance. In the case of multiple eavesdroppers, each of the eavesdroppers can receive the IUI corrupted by the RB vector, indicating that the strategy is still applicable to scenarios with multiple eavesdroppers.

{\bf Creating Strong IUI at Untrusted Users:} To combat internal eavesdropping in NOMA, it is beneficial to create strong IUI intentionally at the untrusted users \cite{CaoYang_WCM2019}. Assume that in a NOMA network, only one specific user requests a secret signal (this user is called a private user), and other users try to eavesdrop the secret signal, i.e., they are untrusted users. To protect the secret signal transmission, the signal beams for each user are appropriately designed, such that the received power of the secret signal is the highest at the private user but the lowest at the other untrusted users. Thus, when the private user decodes the secret signal, it can achieve a high data rate since it receives the secret signal with the highest power. Nevertheless, the secret signal can only be retrieved in the last SIC stage at the untrusted users. Recall that each untrusted user can partially subtract the signals of other untrusted users whose channel gains are stronger, and should leave the remaining signals as IUI when it decodes the secret signal. Having an extremely low secret signal power but a strong IUI power, the eavesdropping rates at the untrusted users are reduced. Therefore, a sufficiently large secrecy sum rate for NOMA can be achieved.

All the previous IUI engineering strategies assume perfect CSI, which, however, is not always the case. In fact, the acquisition of perfect CSI is a non-trivial task. Due to fading, CSI feedback latency, and channel estimation errors, the obtained CSI may be imperfect. This will lead to an IUI design mismatch, e.g., generating strong IUI at the desired user and weak IUI at the eavesdropper, or time-varying IUI at both the desired user and eavesdropper. Thus, robust IUI engineering with imperfect CSI is worth exploring.

A brief summary/comparison of the above-mentioned interference engineering strategies for secure NOMA networks is presented in Table~\ref{table-1}.

\begin{table}[t]
  \centering
  \renewcommand{\arraystretch}{1.0}
  \caption{Interference engineering strategies for secure NOMA communications}\label{table-1}
  \begin{tabular}{|l|p{5.9cm}|p{7cm}|}
  \hline
  \bfseries Strategy & \bfseries Principle & \bfseries Characteristics\\
  \hline\hline
  \multirow{4}*{AN engineering} & Null out the AN in the legitimate channels for secure beamforming; Eliminate the AN at legitimate users by self-interference cancellation for cooperative jamming & Simple but with restrictions on the number of antennas or the served users for secure beamforming; Trade-off between performance and complexity for cooperative jamming \\
  \hline
  \multirow{2}*{SJS engineering} & Enable the users to decode/cancel the SJS but disable the eavesdropper to retrieve the SJS & Wide applicability but need a precise transmission rate or transmit power control for the SJS \\
  \hline
  \multirow{2}*{IUI engineering} & Exploit the IUI as useful jamming or random noise to confound the eavesdropper & Cost-effective but need extra effort in managing interference by user scheduling and beamforming \\
  \hline
  \end{tabular}
\end{table}

\section{Interference Engineering in Advanced NOMA Networks}
\label{sec:chanllenges}

Preliminary works have revealed that big secrecy benefits for NOMA can be drawn from intelligently engineering the network interference. When we integrate NOMA with other advanced wireless communication technologies, more network interference resources can be exploited to enhance security. In the sequel, we propose interference engineering strategies in advanced NOMA networks, and shed light on open research problems and challenges.

\subsection{Interference Engineering in Full-Duplex NOMA}

\begin{figure}[t]
  \centering
  \includegraphics[width=6.5in]{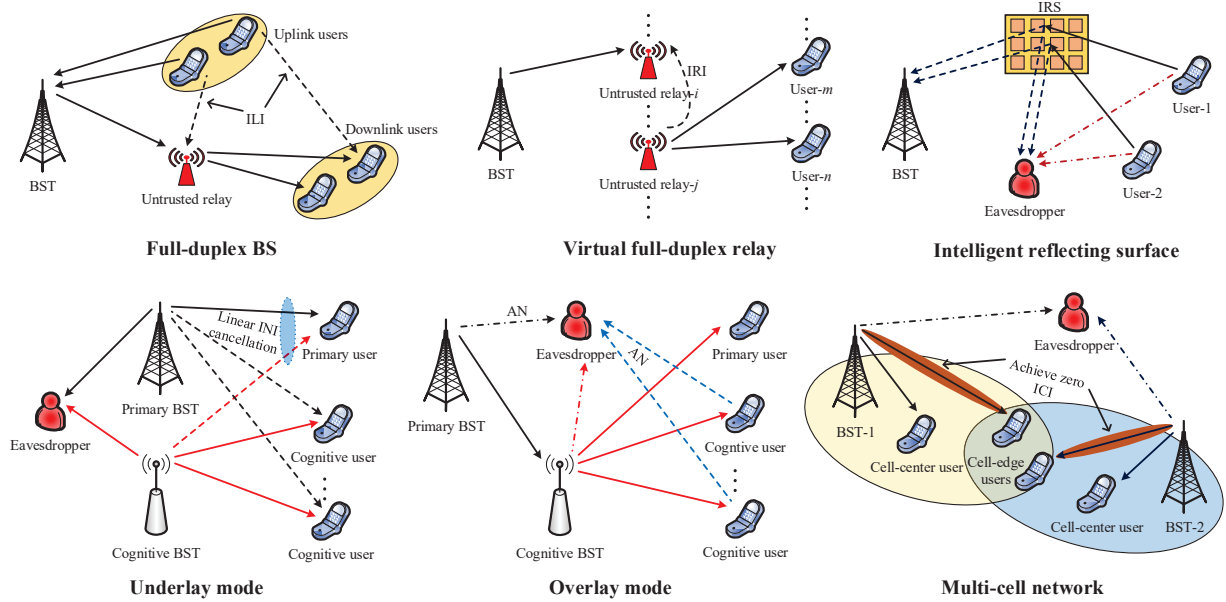}
  \caption{Interference engineering strategies in more advanced NOMA networks.} \label{fig:user-cases}\vspace{-3mm}
\end{figure}

Full-duplex is a cutting-edge technology to improve spectral efficiency by allowing simultaneous information transmission and reception over the same resource block. Thus, the integration of full-duplex techniques into NOMA can bring further performance gains \cite{ZhongCaijun_CL2016}. As shown in the full-duplex BST case of Fig.~\ref{fig:user-cases}, the BST can have uplink and downlink NOMA transmissions simultaneously, while the untrusted relay intercepts signals of both the uplink and downlink. Here, the uplink signals can serve as beneficial interference, i.e., inter-link interference (ILI), to confuse the untrusted relay. On the other hand, the uplink signals are received by the downlink users {\it a priori} and can be used for subsequent self-interference cancellation at the downlink users. To make such cancellation possible, rate adaptation and power allocation are needed to guarantee that the downlink users correctly decode the uplink signals with SIC. Another full-duplex NOMA is illustrated in the virtual full-duplex relay case of Fig.~\ref{fig:user-cases}, where two untrusted relays listen and forward signals successively and the inter-relay interference (IRI) between them can be engineered to combat eavesdropping. In this situation, relay selection and user scheduling are meaningful to maximize the IRI at the untrusted relays while minimizing the IUI at the users, which deserves further investigation.

\subsection{Interference Engineering in Intelligent Reflecting Surface NOMA}

IRS is emerging as a promising technique to enhance the performance of wireless networks, via reconfiguring the radio propagation environment with low-cost passive units integrated on a software-controlled surface \cite{QingqingWu_CM2020}. In particular, the fundamental operating principle of IRS provides an opportunity to engineer the IUI in NOMA constructively at the legitimate receiver but destructively at the eavesdropper. As illustrated in the IRS case of Fig.~\ref{fig:user-cases}, two users apply uplink NOMA to transmit to the BST with the aid of an IRS. By carefully managing the amplitude and/or phase of the IRS, the direct-path and reflected-path signals of user-1 and user-2 can be added at the eavesdropper to create an overlapped signal constellation, which entails a high error rate for the eavesdropper's signal detection and thereby significantly degrades the eavesdropping capability. The signal constellations at the BST can be non-overlapped, and the BST is able to reliably decode the signals. To achieve the above objectives, an optimization framework of transmit power control at the users and passive beamforming design at the IRS are needed, which represent interesting future research opportunities.

\subsection{Interference Engineering in Cognitive Radio NOMA}

Cognitive radio can be integrated with NOMA in a constructive way to realize better spectrum utilization \cite{Lu_CM2018}. In the underlay mode of Fig.~\ref{fig:user-cases}, the inter-network interference (INI) between the simultaneous primary and cognitive transmissions can be exploited as a source of jamming to the eavesdropper. The primary BST transmits combined $s_{\text p}$ and $w_{\text p}s_{\text c}$ and the cognitive BST transmits $w_{\text c}s_{\text c}$, where $s_{\text p}$ denotes the primary user signal, $s_{\text c}$ denotes the superimposed signal for the cognitive users, and $w_{\text p}$ and $w_{\text c}$ denote the weighting coefficients. The designs of $w_{\text p}$ and $w_{\text c}$ are based on the channel gains of the primary and cognitive users, and they facilitate linear INI cancellation of $s_{\text c}$ at the primary user. The cognitive user combines the signals from both BSTs for its information decoding. However, the eavesdropper does not know $w_{\text p}$, $w_{\text c}$ and cannot do SIC to eliminate the INI, such that its capability is degraded. Precise CSI is required in the above design, and efficient channel estimation methods should be devised. In the overlay mode of Fig.~\ref{fig:user-cases}, cooperative jamming via primary and cognitive users can be performed for secrecy, which is explained as follows. First consider that global CSI is available and AN is coherently generated among the cognitive users in a secure fashion (e.g., by the seed of the random noise generator). In the first phase, the cognitive users cooperatively transmit the AN via distributed beamforming when the primary BST is transmitting $s_{\text p}$. In the second phase, the cognitive BST forwards $s_{\text p}$ and $s_{\text c}$ via NOMA to all the users, and at the same time the primary BST injects the AN. The AN in both phases is used to intentionally confuse the eavesdropper. However, it may be hard to have global CSI and precise AN coordination. Accordingly, the cognitive users may transmit independent AN in an uncoordinated way. The independent AN is received by both the cognitive BST and eavesdropper, and thus, it is important to optimize the AN transmit power for striking a balance between degrading the capability of the eavesdropper and affecting the reception quality of the cognitive BST.

\subsection{Interference Engineering in Multi-Cell NOMA}

With the emergence of massive numbers of IoT devices, transceivers are densely deployed and multi-cell NOMA is becoming an attractive candidate for future wireless networks. Contrary to popular belief, the inter-cell interference (ICI) in multi-cell NOMA networks is not detrimental but useful to jam the eavesdropper and improve the secrecy performance. An example of ICI engineering is shown in the multi-cell network case of Fig.~\ref{fig:user-cases}, where two BSTs in different cells jointly optimize their beams to simultaneously guarantee the reliability of the cell-edge users and the security of the overall network. Advanced interference alignment based coordinated beamforming can be applied to achieve zero ICI at the cell-edge users. Since each BST's beam is designed according to the channels between the BST and its serving users, it is not possible for the eavesdropper to cancel the ICI. In such a way, the eavesdropper has to treat the ICI as additional noise to decode the signals, which decreases the eavesdropping rate. For scenarios with multiple users, power allocation and user pairing are crucial to minimize the effect of the ICI to improve the secrecy sum rate. Therefore, it is necessary to investigate a joint design of optimal beamforming, power allocation, and user pairing to enhance the network security.

\section{Concluding Remarks}
\label{sec:conclusion}

In this article, we presented a new view on the benefits of network interference and investigated interference engineering strategies for secure NOMA communications. Specifically, intelligent exploitations of AN, SJS, and IUI against eavesdroppers and/or untrusted users/relays were reviewed. Furthermore, several interference engineering strategies in more advanced full-duplex NOMA, IRS NOMA, cognitive radio NOMA, and multi-cell NOMA networks were proposed, and the corresponding open research problems and challenges were also outlined.

\section*{Acknowledgment}

This work was partially supported by the National Natural Science Foundation of China under Grants 61901313 and 61771366; the Open Research Fund of National Mobile Communications Research Laboratory, Southeast University, under Grant 2020D07; the Natural Science Basic Research Plan of Shaanxi Province under Grants 2020JQ-306 and 2019JQ-197; the China Postdoctoral Science Foundation under Grants BX20190264 and 2019M650258; the US National Science Foundation under Grant NSF-EARS 1547452.

\begin{IEEEbiographynophoto}{Lu Lv}
[M'20] (lulv@xidian.edu.cn) received the Ph.D. degree in communication and information systems from Xidian University, China, in 2018. He is currently an assistant professor with the State Key Laboratory of Integrated Services Networks, Xidian University, and also an academic visitor with the National Mobile Communications Research Laboratory, Southeast University. His research interests include non-orthogonal multiple access, physical layer security, and intelligent reflecting surface.
\end{IEEEbiographynophoto}
\vspace{-20mm}

\begin{IEEEbiographynophoto}{Hai Jiang}
[SM'15] (hai1@ualberta.ca) received the Ph.D. degree in electrical engineering from the University of Waterloo, Waterloo, Ontario, Canada, in 2006. He is currently a professor with the Department of Electrical and Computer Engineering, University of Alberta, Canada. His research interests include radio resource management, cognitive radio networking, and cooperative communications.
\end{IEEEbiographynophoto}
\vspace{-20mm}

\begin{IEEEbiographynophoto}{Zhiguo Ding}
[F'20] (zhiguo.ding@manchester.ac.uk) received the Ph.D. degree in electrical engineering from Imperial College London in 2005. He is currently a chair professor at The University of Manchester, UK. His research interests include 5G communications, MIMO and relaying networks, and energy harvesting. He serves as an Editor for several journals including {\it IEEE Transactions on Communications} and {\it IEEE Transactions on Vehicular Technology}.
\end{IEEEbiographynophoto}
\vspace{-20mm}

\begin{IEEEbiographynophoto}{Qiang Ye}
[M'06] (qye@cs.dal.ca) received the Ph.D. degree in computer science from University of Alberta, Canada. He is currently an associate professor with the Faculty of Computer Science, Dalhousie University, Canada. His research interests include mobile wireless networks, network security, and cloud computing.
\end{IEEEbiographynophoto}
\vspace{-20mm}

\begin{IEEEbiographynophoto}{Naofal Al-Dhahir}
[F'07] (aldhahir@utdallas.edu) received the Ph.D. degree in electrical engineering from Stanford University. He is currently an Erik Jonsson Distinguished Professor at UT-Dallas. He is a co-inventor of 43 issued U.S. patents, co-author of over 450 papers, and a co-recipient of 4 IEEE best paper awards, including the 2006 IEEE Donald G. Fink Award. He served as the Editor-in-Chief of the {\it IEEE Transactions on Communications} from 2016 to 2019.
\end{IEEEbiographynophoto}
\vspace{-20mm}

\begin{IEEEbiographynophoto}{Jian Chen}
[M'14] (jianchen@mail.xidian.edu.cn) received the Ph.D. degree in communication and information systems from Xidian University in 2005. He is currently a professor with the State Key Laboratory of Integrated Services Networks, Xidian University. His research interests include cognitive radio, wireless sensor networks, and heterogeneous networks.
\end{IEEEbiographynophoto}

\end{document}